\documentclass[10pt,journal,compsoc]{IEEEtran}
\renewcommand{\baselinestretch}{0.98}
\ifCLASSOPTIONcompsoc
  \usepackage[nocompress]{cite}
\else
  \usepackage{cite}
\fi

\usepackage{graphicx}
\usepackage{comment}
\usepackage{amsmath,amssymb} 
\usepackage{color}

\usepackage{multirow}
\usepackage{multicol}
\usepackage{floatrow}
\usepackage[noend]{algpseudocode}
\usepackage{algorithmicx,algorithm}
\usepackage{bbding}

\usepackage{tikz,xcolor,hyperref}
\usepackage{subfigure}
\usepackage{booktabs}
\usepackage{makecell}
\usepackage{xspace}

\definecolor{lime}{HTML}{A6CE39}
\DeclareRobustCommand{\orcidicon}{%
    \begin{tikzpicture}
    \draw[lime, fill=lime] (0,0) 
    circle [radius=0.16] 
    node[white] {{\fontfamily{qag}\selectfont \tiny ID}};    \draw[white, fill=white] (-0.0625,0.095) 
    circle [radius=0.007];    \end{tikzpicture}
    \hspace{-2mm}}
\foreach \x in {A, ..., Z}{%
    \expandafter\xdef\csname orcid\x\endcsname{\noexpand\href{https://orcid.org/\csname orcidauthor\x\endcsname}{\noexpand\orcidicon}}
    }

\begin{document}
%
\title{SkinGPT-4: An Interactive Dermatology Diagnostic System with Visual Large Language Model}
%
%
%
%

\author{Juexiao~Zhou$^{1,2,\#}$, Xiaonan He$^{3,\#,*}$, Liyuan Sun$^{4,\#}$, Jiannan Xu$^{4}$, Xiuying Chen$^{1,2}$, Yuetan Chu$^{1,2}$, Longxi Zhou$^{1,2}$, Xingyu Liao$^{1,2}$, Bin Zhang$^{1,2}$, Xin~Gao$^{1,2,*}$
\thanks{
$^1$Computer Science Program, Computer, Electrical and Mathematical Sciences and Engineering Division, King Abdullah University of Science and Technology (KAUST), Thuwal 23955-6900, Kingdom of Saudi Arabia\\
$^2$Computational Bioscience Research Center, King Abdullah University of Science and Technology (KAUST), Thuwal 23955-6900, Kingdom of Saudi Arabia\\
$^3$Emergency Critical Care Center, Beijing AnZhen Hospital, Affiliated to Capital Medical University, Beijing 100029, China\\
$^4$Department of Dermatology, Beijing AnZhen Hospital, Affiliated to Capital Medical University, Beijing 100029, China\\
$^\#$These authors contributed equally.\\
$^*$Corresponding author. e-mail: xin.gao@kaust.edu.sa\\
}}

%
%

\markboth{}%
%
\\
\IEEEtitleabstractindextext{%
\begin{abstract}
Skin and subcutaneous diseases rank high among the leading contributors to the global burden of nonfatal diseases, impacting a considerable portion of the population. Nonetheless, the field of dermatology diagnosis faces three significant hurdles. Firstly, there is a shortage of dermatologists accessible to diagnose patients, particularly in rural regions. Secondly, accurately interpreting skin disease images poses a considerable challenge. Lastly, generating patient-friendly diagnostic reports is usually a time-consuming and labor-intensive task for dermatologists. To tackle these challenges, we present SkinGPT-4, which is the world's first interactive dermatology diagnostic system powered by an advanced visual large language model. SkinGPT-4 leverages a fine-tuned version of MiniGPT-4, trained on an extensive collection of skin disease images (comprising 52,929 publicly available and proprietary images) along with clinical concepts and doctors' notes. We designed a two-step training process to allow SkinGPT-4 to express medical features in skin disease images with natural language and make accurate diagnoses of the types of skin diseases. With SkinGPT-4, users could upload their own skin photos for diagnosis, and the system could autonomously evaluate the images, identifies the characteristics and categories of the skin conditions, performs in-depth analysis, and provides interactive treatment recommendations. Meanwhile, SkinGPT-4's local deployment capability and commitment to user privacy also render it an appealing choice for patients in search of a dependable and precise diagnosis of their skin ailments. To demonstrate the robustness of SkinGPT-4, we conducted quantitative evaluations on 150 real-life cases, which were independently reviewed by certified dermatologists, and showed that SkinGPT-4 could provide accurate diagnoses of skin diseases. Though SkinGPT-4 is not a substitute for doctors, it could enhance users' comprehension of their medical conditions, facilitate improve communication between patients and doctors, expedite the diagnostic process for dermatologists, and potentially promote human-centred care and healthcare equity in underdeveloped areas.

\end{abstract}

}

\maketitle

\IEEEdisplaynontitleabstractindextext

%
\IEEEpeerreviewmaketitle

\section{Introduction}
Skin and subcutaneous diseases rank as the fourth major cause of nonfatal disease burden worldwide, affecting a considerable proportion of individuals, with a prevalence ranging from 30\% to 70\% across all ages and regions\cite{hay2014global}. However, dermatologists are consistently in short supply, particularly in rural areas, and consultation costs are on the rise\cite{feng2018comparison, resneck2004dermatology, liu2020deep}. As a result, the responsibility of diagnosis often falls on non-specialists such as primary care physicians, nurse practitioners, and physician assistants, which may have limited knowledge and training \cite{seth2017global} and low accuracy on diagnosis\cite{federman1999comparison, moreno2007prospective}. The use of store-and-forward teledermatology has become dramatically popular in order to expand the range of services available to medical professionals\cite{yim2018teledermatology}, which involves transmitting digital images of the affected skin area (usually taken using a digital camera or smartphone)\cite{kshirsagar2022deep} and other relevant medical information from users to dermatologists. Then, the dermatologist reviews the case remotely and advises on diagnosis, workup, treatment, and follow-up recommendations\cite{martora2022patient, lopez2022teledermatology}. Nonetheless, the field of dermatology diagnosis faces three significant hurdles\cite{lakdawala2022workforce}. Firstly, there is a shortage of dermatologists accessible to diagnose patients, particularly in rural regions. Secondly, accurately interpreting skin disease images poses a considerable challenge. Lastly, generating patient-friendly diagnostic reports is usually a time-consuming and labor-intensive task for dermatologists \cite{liu2020deep, pious2022review}.

Advancements in technology have led to the development of various tools and techniques to aid dermatologists in their diagnosis\cite{pious2022review, puri2022deep, reshma2023review}. For example, the development of artificial intelligence tools to aid in the diagnosis of skin disorders from images has been made possible by recent advancements in deep learning\cite{han2020augmented, popescu2022new}, such as skin cancer classification\cite{esteva2017dermatologist, han2018classification, haenssle2018man, marchetti2018results, brinker2019comparing, yap2018multimodal, aggarwal2019data, tschandl2019expert, han2020keratinocytic, jones2022artificial}, dermatopathology\cite{hekler2019pathologist, jiang2020recognizing, hekler2019deep}, predicting novel risk factors or epidemiology\cite{roffman2018predicting, lott2018population}, identifying onychomycosis\cite{han2018deep}, quantifying alopecia areata\cite{bernardis2018quantifying}, classify skin lesions from mpox virus infection\cite{thieme2023deep}, and so on\cite{liu2020deep}. Among these, most studies have predominantly concentrated on identifying skin lesions through dermoscopic images\cite{cruz2013deep, yuan2017automatic, tschandl2019comparison}. However, dermatoscopy is often not readily available outside of dermatology clinics. Some studies have explored the use of clinical photographs of skin cancer\cite{esteva2017dermatologist}, onychomycosis\cite{han2018deep}, and skin lesions on educational websites\cite{sun2016benchmark}. Nevertheless, those methods are tailored for particular diagnostic objectives as classification tasks and their approach still requires further analysis by dermatologists to issue reports and make clinical decisions. Those methods are unable to automatically generate detailed reports in natural language and allow interactive dialogues with patients. At present, there are no such diagnostic systems available for users to self-diagnose skin conditions by submitting images that can automatically and interactively analyze and generate easy-to-understand text reports.

Over the past few months, the field of large language models (LLMs) has seen significant advancements\cite{kung2023performance, sallam2023chatgpt}, offering remarkable language comprehension abilities and the potential to perform complex linguistic tasks. One of the most anticipated models is GPT-4\cite{bubeck2023sparks}, which is a large-scale multimodal model that has demonstrated exceptional capabilities, such as generating accurate and detailed image descriptions, providing explanations for atypical visual occurrences, constructing websites based on handwritten textual descriptions, and even acting as family doctors\cite{lee2023benefits}. Despite these remarkable advancements, some features of GPT-4 are still not accessible to the public and are closed-source. Users need to pay and use some features through API. As an accessible alternative, ChatGPT, which is also developed by OpenAI, has demonstrated the potential to assist in disease diagnosis through conversation with patients\cite{balas2023conversational, mijwil2023chatgpt, sinha2023applicability, sinha2023applicability, ufuk2023role, hu2023advancing, vaishya2023chatgpt}. By leveraging its advanced natural language processing capabilities, ChatGPT could interpret symptoms and medical history provided by patients and make suggestions for potential diagnoses or referrals to appropriate dermatological specialists\cite{beltrami2023consulting}. However, ChatGPT currently only allows text input and does not support direct image input for diagnosis, which limits its availability for dermatological diagnosis. 

\begin{figure*}[!htb]
    \centering
    \includegraphics[width=1\linewidth]{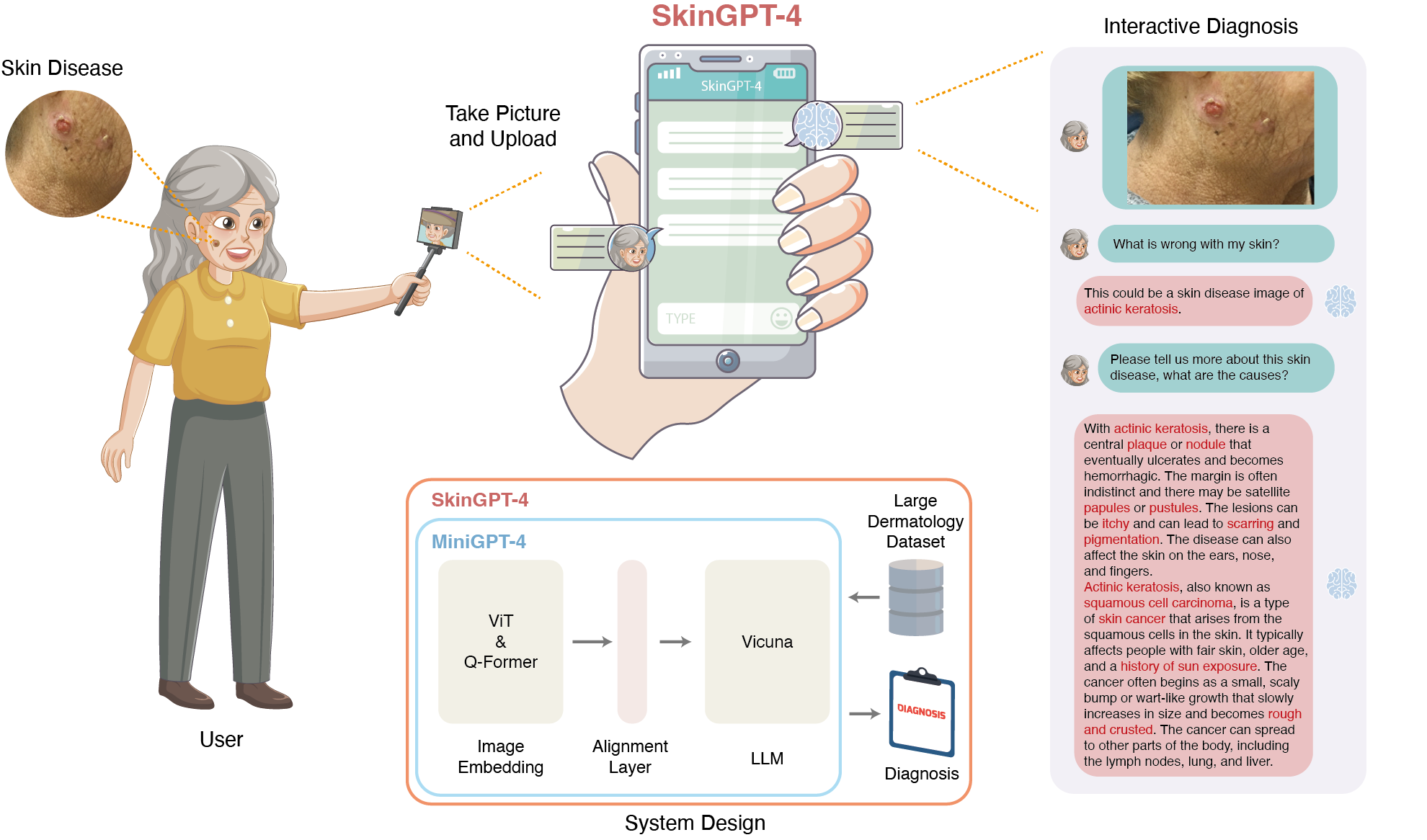}
    \caption{Illustration of SkinGPT-4. SkinGPT-4 incorporates a fine-tuned version of MiniGPT-4 on a vast collection (52,929) of both public and in-house skin disease images, accompanied by clinical concepts and doctors' notes. With SkinGPT-4, users could upload their own skin photos for diagnosis, and SkinGPT-4 could autonomously determine the characteristics and categories of skin conditions, perform analysis, provide treatment recommendations, and allow interactive diagnosis. On the right is an example of interactive diagnosis.}
    \label{fig_SkinGPT-4}
\end{figure*}

\begin{figure*}[!htb]
    \centering
    \includegraphics[width=1\linewidth]{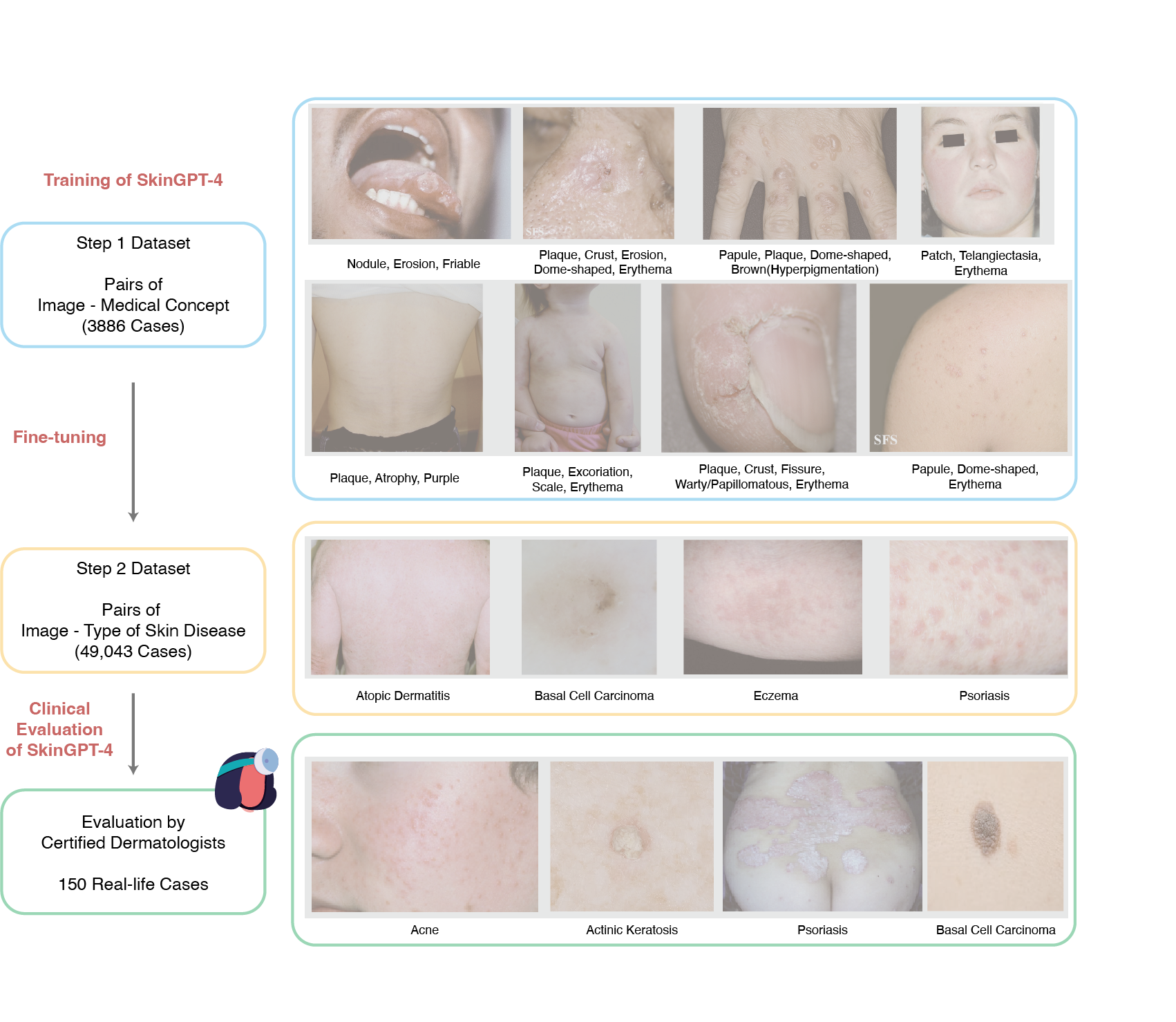}
    \caption{Illustration of our datasets for two-step training of SkinGPT-4. The notes below each image indicate clinical concepts and types of skin diseases. In addition, we have detailed descriptions from the certified dermatologists for images in the step 2 dataset. To avoid causing discomfort, we used a translucent grey box to obscure the displayed skin disease images.}
    \label{fig_dataset}
\end{figure*}

\begin{figure*}[!htb]
    \centering
    \includegraphics[width=1\linewidth]{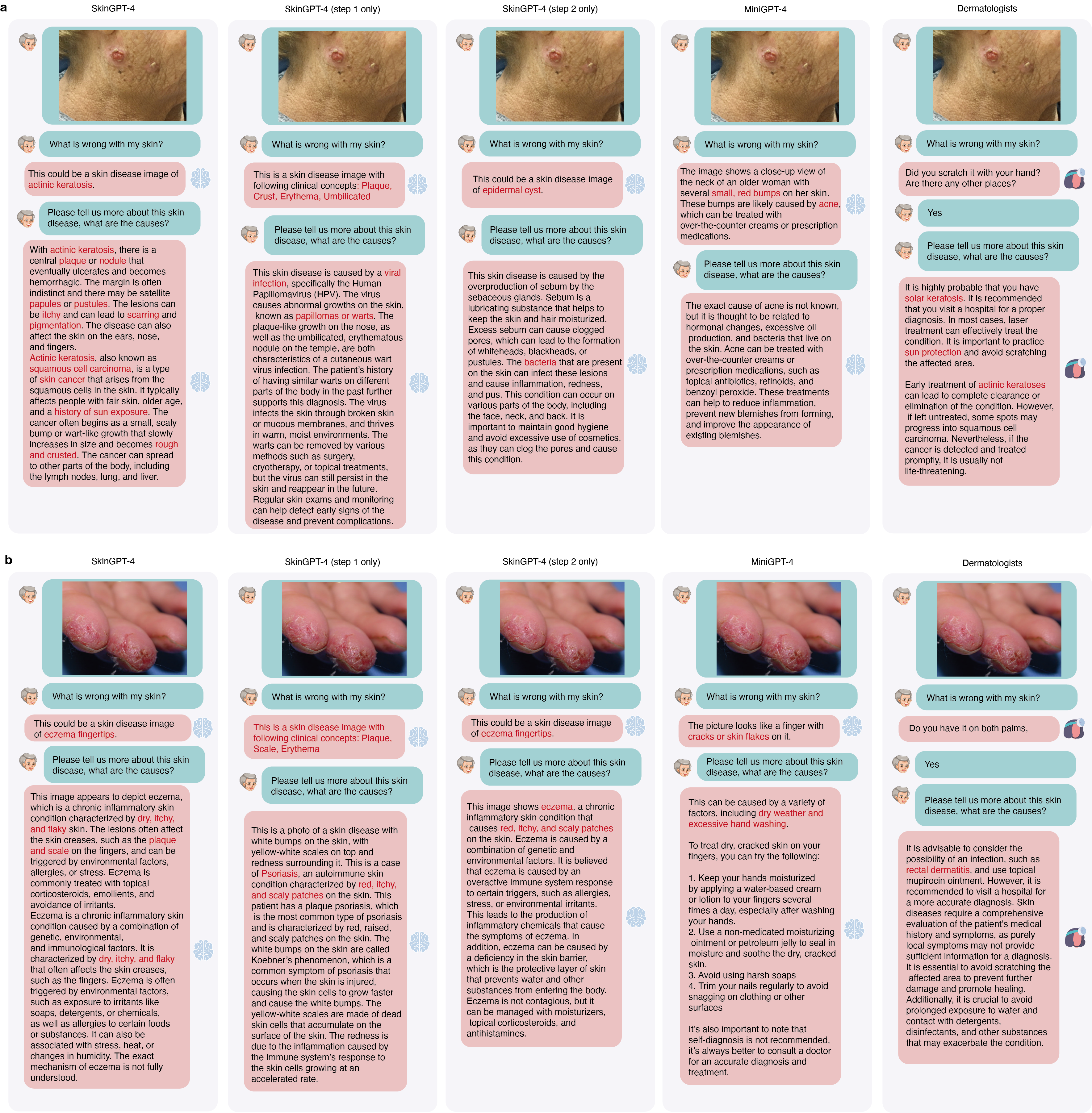}
    \caption{Diagnosis generated by SkinGPT-4, SkinGPT-4 (step 1 only), SkinGPT-4 (step 2 only), MiniGPT-4 and Dermatologists. \textbf{a}. A case of actinic keratosis. \textbf{b}. A case of eczema fingertips.}
    \label{fig_comparison}
\end{figure*}

The idea of providing skin images directly for automatic dermatological diagnosis and generating text reports is exciting because it could greatly help solve the three aforementioned challenges in the field of dermatology diagnosis. However, there exists no method to accomplish this at present. But in related areas, ChatCAD\cite{wang2023chatcad} is one of the most advanced approaches that designed various networks to take X-rays, CT scans, and MRIs images to generate diverse outputs, which are then transformed into text descriptions. These descriptions are combined as inputs to ChatGPT to generate a condensed report and offer interactive explanations and medical recommendations based on the given image. However, their proposed vision-text models were limited to certain tasks. Meanwhile, for ChatCAD, users need to use ChatGPT's API to upload text descriptions, which could raise data privacy issues\cite{li2023multi, lund2023information, sallam2023chatgpt} as both medical images and text descriptions contain a lot of patients' private information\cite{rajpurkar2022ai, zhou2022ppml, zhou2023personalized, zhou2023audit}. To address those issues, MiniGPT-4 \cite{zhu2022minigpt4} is the first open-source method that allows users to deploy locally to interface images with state-of-the-art LLMs and interact using natural language without the need to fine-tune both pre-trained large models but only a small alignment layer. MiniGPT-4 aims to combine the power of a large language model with visual information obtained from a pre-trained vision encoder. To achieve this, the model uses Vicuna\cite{Vicuna} as its language decoder, which is built on top of LLaMA\cite{touvron2023llama} and is capable of performing complex linguistic tasks. To process visual information, the same visual encoder used in BLIP-2\cite{li2023blip} is employed, which consists of a ViT\cite{fang2022eva} backbone combined with a pre-trained Q-Former. Both the language and vision models are open-source. To bridge the gap between the visual encoder and the language model, MiniGPT-4 utilizes a linear projection layer. However, MiniGPT-4 is trained on the combined dataset of Conceptual Caption\cite{Conceptual_captions}, SBU\cite{ordonez2011im2text}, and LAION\cite{schuhmann2021laion}, which are irrelevant to medical images, especially dermatological images. Therefore, it is still challenging to directly apply MiniGPT-4 to specific domains such as formal dermatology diagnosis.

\begin{figure*}[!htb]
    \centering
    \includegraphics[width=1\linewidth]{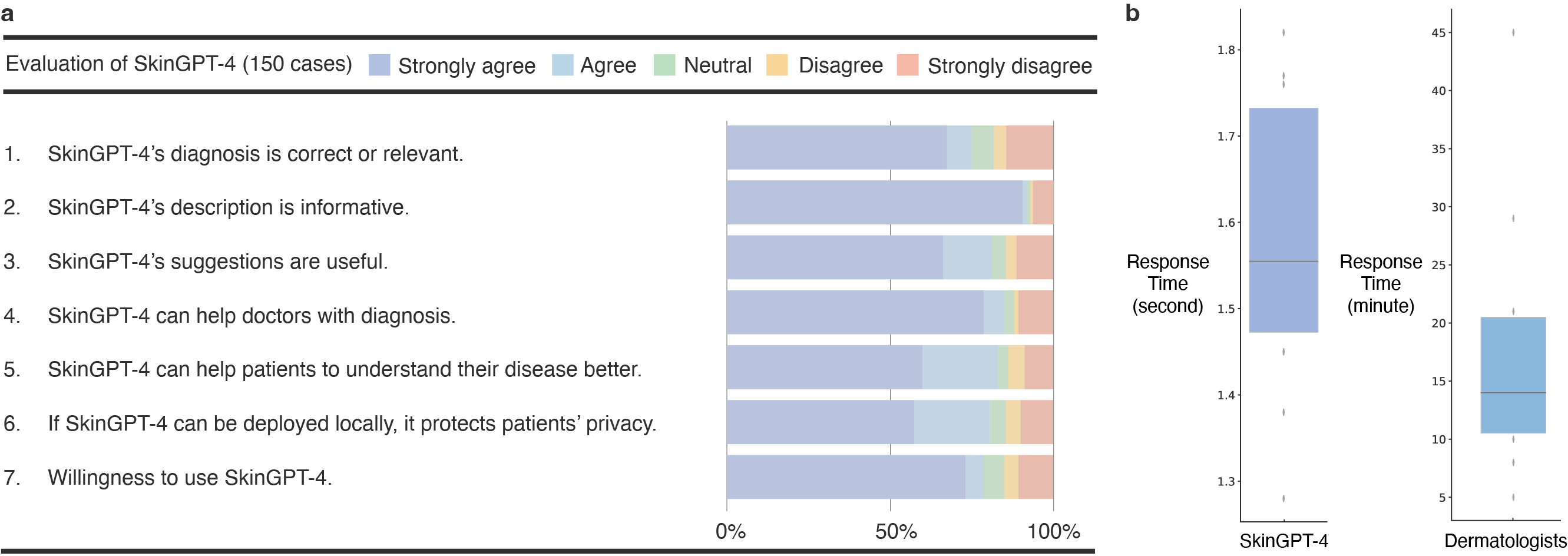}
    \caption{Clinical evaluation of SkinGPT-4 by certified offline and online dermatologists. \textbf{a}. Questionnaire-based assessment of SkinGPT-4 by offline dermatologists. \textbf{b}. Response time of SkinGPT-4 compared to consulting dermatologists online.}
    \label{fig_response_time}
\end{figure*}

Here, we propose SkinGPT-4, the world's first dermatology diagnostic system powered by an advanced vision-based large language model (\textbf{Figure} \ref{fig_SkinGPT-4}). SkinGPT-4 leverages a fine-tuned version of MiniGPT-4, trained on an extensive collection of skin disease images (comprising 52,929 publicly available and proprietary images) along with clinical concepts and doctors' notes. We designed a two-step training process to develop SkinGPT-4 as shown in \textbf{Figure} \ref{fig_dataset}. In the initial step, SkinGPT-4 aligns visual and textual clinical concepts, enabling it to recognize medical features within skin disease images and express those medical features with natural language. In the subsequent step, SkinGPT-4 learns to accurately diagnoses the specific types of skin diseases. This comprehensive training methodology ensures the system's proficiency in analyzing and classifying various skin conditions. With SkinGPT-4, users have the ability to upload their own skin photos for diagnosis. The system autonomously evaluates the images, identifies the characteristics and categories of the skin conditions, performs in-depth analysis, and provides interactive treatment recommendations (\textbf{Figure} \ref{fig_comparison}). Moreover, SkinGPT-4's localized deployment capability and a strong commitment to user privacy make it a trustworthy and precise diagnostic tool for patients seeking reliable assessments of their skin ailments. Meanwhile, we showed that SkinGPT-4 could empower patients to gain a clearer understanding of their symptoms, diagnosis, and treatment plans, which could help patients engage in more effective and economical consultations with dermatologists. With SkinGPT-4, patients can have more informed conversations with their doctors, leading to better treatment outcomes and a higher level of satisfaction. To demonstrate the robustness of SkinGPT-4, we conducted quantitative evaluations on 150 real-life cases, which were independently reviewed by certified dermatologists (\textbf{Figure} \ref{fig_response_time} and \textbf{Supplementary information}). The results showed that SkinGPT-4 consistently provided accurate diagnoses of skin diseases. It is important to note that while SkinGPT-4 is not a substitute for medical professionals, it greatly enhances users' understanding of their medical conditions, facilitates improved communication between patients and doctors, expedites the diagnostic process for dermatologists, and has the potential to advance human-centred care and healthcare equity, particularly in underdeveloped regions\cite{preiksaitis2023chatgpt}. In summary, SkinGPT-4 represents a significant leap forward in the field of dermatology diagnosis in the era of large language models.

\section{Results}

\subsection{The Overall Design of SkinGPT-4}
SkinGPT-4 is an interactive system designed to provide a natural language-based diagnosis of skin disease images as shown in Figure \ref{fig_SkinGPT-4}. The process commences when the user uploads a skin image, which undergoes encoding by the Vision Transformer (VIT) and Q-Transformer models to comprehend its contents. The VIT model partitions the image into smaller patches and extracts vital features like edges, textures, and shapes. After that, the Q-Transformer model generates an embedding of the image based on the features identified by the VIT model, which is done by using a transformer-based architecture that allows the model to consider the context of the image. The alignment layer facilitates the synchronization of visual information and natural language, and the Vicuna component generates the text-based diagnosis. SkinGPT-4 is fine-tuned on MiniGPT-4 using large skin disease images along with clinical concepts and doctors' notes to allow for interactive dermatological diagnosis. The system could provide an interactive and user-friendly way to help users self-diagnose skin diseases.

\subsection{Interactive, Informative and Understandable Dermatology Diagnosis of SkinGPT-4}
SkinGPT-4 brings forth a multitude of advantages for both patients and dermatologists. One notable benefit lies in its utilization of comprehensive and trustworthy medical knowledge specifically tailored to skin diseases. This empowers SkinGPT-4 to deliver interactive diagnoses, explanations, and recommendations for skin diseases (\textbf{Supplementary Video}), which presents a challenge for MiniGPT-4. Unlike MiniGPT-4, which lacks training with pertinent medical knowledge and domain-specific adaptation, SkinGPT-4 overcomes this limitation, enhancing its proficiency in the dermatological domain. To demonstrate the advantage of SkinGPT-4 over MiniGPT-4, we presented two real-life examples of interactive diagnosis as shown in Figure \ref{fig_comparison}. In Figure \ref{fig_comparison}a, an image is presented of an elderly with actinic keratosis on her face. In Figure \ref{fig_comparison}b, an image is provided of a patient with eczema fingertips.

For the actinic keratosis case (Figure \ref{fig_comparison}a), MiniGPT-4 identified features like small and red bumps, and incorrectly diagnosed the skin disease as acne, while SkinGPT-4 identified features like plaque, nodules, pustules, and scarring, and diagnosed the skin disease as actinic keratosis, which is a common skin condition caused by prolonged exposure to the sun's ultraviolet (UV) rays\cite{fu2003actinic}. During the interactive dialogue, SkinGPT-4 also suggested the cause of the skin disease to be sun exposure, which was also verified as correct by the certified dermatologist. For the example of eczema fingertips case (Figure \ref{fig_comparison}b), MiniGPT-4 identified some features like cracks and skin flakes, missed the type of the skin disease, and diagnosed the cause of the skin disease to be dry weather and excessive hand washing. In comparison, SkinGPT-4 identified either the features of the skin disease as dry itchy and flaky skin, and diagnosed the type of the skin disease to be eczema fingertips, which was also verified by certified dermatologists.

In summary, the absence of dermatological knowledge and domain-specific adaptation poses a significant challenge for MiniGPT-4 in achieving accurate dermatological diagnoses. Contrastingly, SkinGPT-4 successfully and accurately identified the characteristics of the skin diseases displayed in the images. It not only suggested potential disease types but also provided recommendations for potential treatments. This further highlights that domain-specific adaption is crucial for SkinGPT-4 to work for the dermatological diagnosis.

\subsection{SkinGPT-4 Masters Medical Features to Improve Diagnosis with the Two-step Training}
To further illustrate the capability of SkinGPT-4 in enhancing dermatological diagnosis through learning medical features in skin disease images, we conducted ablation studies, as depicted in Figure \ref{fig_comparison} by training SkinGPT-4 using either solely the step 1 dataset or solely the step 2 dataset. As specified in \textbf{Method} and illustrated in Figure \ref{fig_dataset}, we designed a two-step training process for SkinGPT-4. Initially, we utilized the step 1 dataset to familiarize SkinGPT-4 with the medical features present in dermatological images and allow SkinGPT-4 to express medical features in skin disease images with natural language. Subsequently, we employed the step 2 dataset to train SkinGPT-4 to achieve a more precise diagnosis of disease types.

In the instance of actinic keratosis (Figure \ref{fig_comparison}a), which is a hard case, SkinGPT-4 trained solely on the step 1 dataset demonstrated its proficiency in identifying pertinent medical features such as plaque, crust, erythema, and umbilicated. These precise and comprehensive morphological descriptions accurately captured the characteristics of the skin disease depicted in the image. However, when SkinGPT-4 was exclusively trained on the step 1 dataset, it erroneously diagnosed the skin condition as a viral infection, indicating the importance of incorporating the step 2 dataset for more accurate disease identification. In contrast, when trained solely on the step 2 dataset, SkinGPT-4 failed to capture the accurate morphological descriptions of the skin diseases and instead incorrectly diagnosed it as the result of excessive sebum production. It highlights the necessity of incorporating the step 1 dataset to effectively recognize and comprehend the specific medical features essential for precise dermatological diagnoses. In comparison, SkinGPT-4 with our two-step training simultaneously identified the medical features, such as plaque, nodules, pustules and scarring, and diagnosed the skin disease as actinic keratosis. For simple cases such as the eczema fingertips shown in Figure \ref{fig_comparison}b, SkinGPT-4 could also provide more detailed descriptions of the skin disease image, encompass the medical features and accurately identify the type of skin disease. In conclusion, the two-step training process we have implemented allows SkinGPT-4 to effectively comprehend and master medical features in dermatological images, thereby significantly enhancing the accuracy of diagnoses, which is particularly crucial for hard cases where precise identification of medical features is paramount to accurately determining the type of disease.

\subsection{Clinical Evaluation of SkinGPT-4 by Certified Dermatologists}
To evaluate the reliability and robustness of SkinGPT-4, we conducted a comprehensive study involving a large number of real-life cases (150) and compared its diagnoses with those of certified dermatologists. The results, presented in Table \ref{table_step2_dataset} and Supplementary information, demonstrated that SkinGPT-4 consistently provided accurate diagnoses that were in agreement with those of the certified dermatologists as shown in Figure \ref{fig_response_time}, as well as in all cases detailed in the Supplementary information.

Among the 150 cases, a significant percentage of SkinGPT-4's diagnoses (78.76\%) were evaluated as correct or relevant by certified dermatologists. This evaluation encompassed both strongly agree (73.13\%) and agree (5.63\%). Additionally, SkinGPT-4's responses regarding the causes of the disease and potential treatments were considered informative (80.63\%) and useful (83.13\%) by the doctors. Furthermore, SkinGPT-4 proved to be a valuable tool for doctors in the diagnosis process (85\%) and for patients in gaining a better understanding of their diseases (81.25\%). The capability of SkinGPT-4 to support local deployment, ensuring user privacy, garnered high agreement (91.88\%), further enhancing the willingness to utilize SkinGPT-4 (75\%).

Overall, the study demonstrated that SkinGPT-4 delivers reliable diagnoses, aids doctors in the diagnostic process, facilitates patient understanding, and prioritizes user privacy, making it a valuable asset in the field of dermatology.

\subsection{SkinGPT-4 Acts as a 24/7 On-call Family Doctor}
In comparison to online consultations with dermatologists, which often entail waiting minutes for a response, SkinGPT-4 offers several advantages. Firstly, it is available 24/7, ensuring constant access to medical advice. Additionally, SkinGPT-4 provides faster response times, typically within seconds, as depicted in Figure \ref{fig_response_time}b, which makes it a swift and convenient option for patients requiring immediate diagnoses outside of regular office hours.

Moreover, SkinGPT-4's ability to offer preliminary diagnoses empowers patients to make informed decisions about seeking in-person medical attention. This feature can help reduce unnecessary visits to the doctor's office, saving patients both time and money. The potential to improve healthcare access is particularly significant in rural areas or regions experiencing a scarcity of dermatologists. In such areas, patients often face lengthy waiting times or must travel considerable distances to see a dermatologist \cite{kanthraj2023twenty}. By leveraging SkinGPT-4, patients can swiftly and conveniently receive preliminary diagnoses, potentially diminishing the need for in-person visits and alleviating the strain on healthcare systems in these underserved regions.

\begin{table}[!htb]
    \centering
        \caption{Characteristics of Step 1 Dataset. It is possible for a single image to have multiple medical concepts at the same time. The total number of samples is 3886.
        }
        \small
    \begin{tabular}{c|c}
\toprule
Clinical Concepts  & Number of Samples \\
\toprule
Erythema &   2139 \\
                  Plaque &   1966 \\
                  Papule &   1169 \\
Brown(Hyperpigmentation) &    759 \\
                   Scale &    686 \\
                   Crust &    497 \\
 White(Hypopigmentation) &    257 \\
                  Yellow &    245 \\
                 Erosion &    200 \\
                  Nodule &    189 \\
                   Ulcer &    154 \\
                 Friable &    153 \\
                   Patch &    149 \\
             Dome-shaped &    146 \\
                 Exudate &    144 \\
                    Scar &    123 \\
                 Pustule &    103 \\
          Telangiectasia &    100 \\
                   Black &     90 \\
                  Purple &     85 \\
                 Atrophy &     69 \\
                   Bulla &     64 \\
             Umbilicated &     49 \\
                 Vesicle &     46 \\
     Warty/Papillomatous &     46 \\
             Excoriation &     46 \\
     Exophytic/Fungating &     42 \\
                 Xerosis &     35 \\
              Induration &     33 \\
                 Fissure &     32 \\
               Sclerosis &     27 \\
            Pedunculated &     26 \\
         Lichenification &     25 \\
                  Comedo &     24 \\
                   Wheal &     21 \\
             Flat topped &     18 \\
             Translucent &     16 \\
                  Macule &     13 \\
                  Salmon &     10 \\
       Purpura/Petechiae &     10 \\
               Acuminate &      8 \\
                    Cyst &      6 \\
                    Blue &      5 \\
                 Abscess &      5 \\
            Poikiloderma &      5 \\
                  Burrow &      5 \\
                    Gray &      5 \\
               Pigmented &      5 \\
\bottomrule
\end{tabular}
    \label{table_step1_dataset}
\end{table}

\begin{table*}[!htb]
    \centering
        \caption{Characteristics of Step 2 Dataset and Clinical Evaluation Dataset.
        }
        \footnotesize
    \begin{tabular}{c|c|c}
\toprule
Major Classes of Skin Disease  & \makecell{Number of Samples\\in Step 2 Dataset} & \makecell{Number of Samples\\in Clinical Evaluation Dataset}\\
\toprule
Acne and Rosacea & 840 & 10 \\
Malignant Lesions (Actinic Keratosis, Basal Cell Carcinoma, etc.) & 8166 & 10 \\
Dermatitis (Atopic Dermatitis, Eczema, Exanthems, Drug Eruptions, Contact Dermatitis, etc.) & 5262 & 10 \\
Bullous Disease & 448 & 10 \\
Bacterial Infections (Cellulitis, Impetigo, etc.) & 228 & 10 \\
Light Diseases (vitiligo, sun damaged skin, etc.) & 568 & 10 \\
Connective Tissue diseases (Lupus, etc.) & 420 & 10 \\
Benign Tumors (Seborrheic Keratoses, etc.)  & 1916 & 10 \\
Melanoma Skin Cancer, Nevi, Moles & 23373 & 10 \\
Fungal Infections (Nail Fungus, Tinea Ringworm, Candidiasis, etc.) & 2340 & 10\\
Psoriasis and Lichen Planus & 3460 & 10\\
Infestations and Bites (Scabies, Lyme Disease, etc.) & 431 & 10\\
Urticaria Hives & 212 & 10\\
Vascular Tumors & 735 & 10\\
Herpes & 405 & 10 \\
Others & 239 & / \\
\toprule
Total & 49043 & 150\\
\bottomrule
\end{tabular}
    \label{table_step2_dataset}
\end{table*}

\section{Methods}
\subsection{Dataset}
Our datasets include two public datasets and our private in-house dataset, where the first public dataset was used for the step 1 training, and the second public dataset and our in-house dataset were used for the step 2 training. 

The first public dataset named SKINCON\cite{daneshjou2022skincon} is the first medical dataset densely annotated by domain experts to provide annotations useful across multiple disease processes. SKINCON is a skin disease dataset densely annotated by dermatologists and it includes 3230 images from the Fitzpatrick 17k skin disease dataset densely annotated with 48 clinical concepts as shown in \textbf{Table} \ref{table_step1_dataset}, 22 of which have at least 50 images representing the concept, and 656 skin disease images from the Diverse Dermatology Images dataset. The 48 clinical concepts proposed by SKINCON include Vesicle, Papule, Macule, Plaque, Abscess, Pustule, Bulla, Patch, Nodule, Ulcer, Crust, Erosion, Excoriation, Atrophy, Exudate, Purpura/Petechiae, Fissure, Induration, Xerosis, Telangiectasia, Scale, Scar, Friable, Sclerosis, Pedunculated, Exophytic/Fungating, Warty/Papillomatous, Dome-shaped, Flat-topped, Brown (Hyperpigmentation), Translucent, White (Hypopigmentation), Purple, Yellow, Black, Erythema, Comedo, Lichenification, Blue, Umbilicated, Poikiloderma, Salmon, Wheal, Acuminate, Burrow, Gray, Pigmented, and Cyst. 

The second public dataset named the Dermnet contains 18,856 images, which are further classified into 15 classes by our board-certified dermatologists, including Acne and Rosacea, Malignant Lesions (Actinic Keratosis, Basal Cell Carcinoma, etc.), Dermatitis (Atopic Dermatitis, Eczema, Exanthems, Drug Eruptions, Contact Dermatitis, etc.), Bullous Disease, Bacterial Infections (Cellulitis, Impetigo, etc.), Light Diseases (vitiligo, sun damaged skin, etc.), Connective Tissue diseases (Lupus, etc.), Benign Tumors (Seborrheic Keratoses, etc.), Melanoma Skin Cancer (Nevi, Moles, etc.), Fungal Infections (Nail Fungus, Tinea Ringworm, Candidiasis, etc.), Psoriasis and Lichen Planus, Infestations and Bites (Scabies, Lyme Disease, etc.), Urticaria Hives, Vascular Tumors, Herpes, and others. 

Our private in-house dataset contains 30,187 pairs of skin disease images and corresponding doctors' descriptions. The complete dataset for step 2 training comprises in total of 49,043 pairs of images and textual descriptions as shown in \textbf{Table} \ref{table_step2_dataset}.

\subsection{The two-step training of SkinGPT-4}
SkinGPT-4 was trained using a vast of skin disease images along with clinical concepts and doctors' notes (Figure \ref{fig_SkinGPT-4}). In the first step, we fine-tuned the pre-trained MiniGPT-4 model using the step 1 training dataset. This dataset consists of paired skin disease images along with corresponding descriptions of clinical concepts. By training SkinGPT-4 on this dataset, we enabled the model to grasp the nuances of clinical concepts specific to skin diseases. 

In the second step, we further refined the model by fine-tuning it using the step 2 dataset, which comprises additional skin images and refined doctors' notes. This iterative training process facilitated the accurate diagnosis of various skin diseases, as SkinGPT-4 incorporated the refined medical insights from the doctors' notes.

By following this two-step fine-tuning approach, SkinGPT-4 attained an enhanced understanding of clinical concepts related to skin diseases and acquired the proficiency to generate accurate diagnoses.

\subsection{Model Training and Resources}
During the training of both steps, the max number of epochs was fixed to 20, the iteration of each epoch was set to 5000, the warmup step was set to 5000, batch size was set to 2, the learning rate was set to 1e-4, and max text length was set to 160. The entire fine-tuning process required approximately 9 hours to complete and utilized two NVIDIA V100 (32GB) GPUs. During inference, only one NVIDIA V100 (32GB) GPU was necessary. SkinGPT-4 was developed using Python 3.7, PyTorch 1.9.1, and CUDA 11.4. For a comprehensive list of dependencies, please refer to our code availability documentation. The training and inference procedures were conducted on a workstation equipped with 252 GB RAM, 112 CPU cores, and two NVIDIA V100 GPUs, which provided the computational resources necessary for efficient model training and inference.

\subsection{Clinical Evaluation of SkinGPT-4}
To assess the reliability and effectiveness of SkinGPT-4, we assembled a dataset comprising 150 real-life cases of various skin diseases as shown in Table \ref{table_step2_dataset}. Interactive diagnosis sessions were conducted with SkinGPT-4, utilizing four specific prompts:

1. Could you describe the skin disease in this image for me?

2. Please provide a paragraph listing additional features you observed in the image.

3. Based on the previous information, please provide a detailed explanation of the cause of this skin disease.

4. What treatment and medication should be recommended for this case?

To conduct the clinical evaluation, certified dermatologists were provided with the same set of four questions and were required to make diagnoses based on the given skin disease images. Meanwhile, the dermatologists also evaluated the reports generated by SkinGPT-4 and assigned scores (strongly agree, agree, neutral, disagree, and strongly disagree) to each item in the evaluation form (Figure \ref{fig_response_time}a), including the following questions:

1. SkinGPT-4’s diagnosis is correct or relevant.

2. SkinGPT-4’s description is informative.

3. SkinGPT-4’s suggestions are useful.

4. SkinGPT-4 can help doctors with diagnosis.

5. SkinGPT-4 can help patients to understand their disease better.

6. If SkinGPT-4 can be deployed locally, it protects patients’ privacy.

7. Willingness to use SkinGPT-4.

In particular, for questions 3 and 5, we further collected the opinions of users of SkinGPT-4, who usually do not have strong background knowledge in dermatology, to show that SkinGPT-4 is friendly to the general users. Those results allowed for a comprehensive evaluation of SkinGPT-4's performance in relation to certified dermatologists and patients.

\section{Conclusion and Discussion}
Our study showcases the promising potential of utilizing visual inputs in LLMs to enhance dermatological diagnosis. With the upcoming release of more advanced LLMs like GPT-4, the accuracy and quality of diagnoses could be further improved. However, it is essential to address potential privacy concerns associated with using LLMs like ChatGPT and GPT-4 as an API, as it requires users to upload their private data. In contrast, SkinGPT-4 offers a solution to this privacy issue. By allowing users to deploy the model locally, the concerns regarding data privacy are effectively resolved. Users have the autonomy to use SkinGPT-4 within the confines of their own system, ensuring the security and confidentiality of their personal information.

During the course of a patient's consultation with a dermatologist, the doctor often asks additional questions to gather crucial information that aids in arriving at a precise diagnosis. In contrast, SkinGPT-4 relies on the information provided by users to assist in the diagnostic process. Additionally, doctors often engage in empathetic interactions with patients, as the emotional connection could contribute to the diagnostic process. Due to these factors, it remains challenging for SkinGPT-4 to fully replace dermatologists at present. However, SkinGPT-4 still holds significant value as a tool for both patients and dermatologists. It can greatly expedite the diagnostic process and enhance the overall service delivery. By leveraging its capabilities, SkinGPT-4 empowers patients to obtain preliminary insights into their skin conditions and aids dermatologists in providing more efficient care. While it may not fully substitute for the expertise and empathetic nature of dermatologists, SkinGPT-4 serves as a valuable complementary resource in the field of dermatological diagnosis.

As LLMs-based applications like SkinGPT-4 continue to evolve and improve with the acquisition of even more reliable medical training data, the potential for significant advancements in online medical services is enormous. SkinGPT-4 could play a critical role in improving access to healthcare and enhancing the quality of medical services for patients worldwide. We will continue our research in this field to further develop and refine this technology.

\section{Acknowledgements}
\noindent
\textbf{Special thanks: }Thanks to Jun Chen, the author of MiniGPT-4 for the discussion of this work.\\

\noindent
\textbf{Funding: }Juexiao Zhou, Xiuying Chen, Yuetan Chu, Longxi Zhou, Xingyu Liao, Bin Zhang, and Xin Gao were supported in part by grants from the Office of Research Administration (ORA) at King Abdullah University of Science and Technology (KAUST) under award number FCC/1/1976-44-01, FCC/1/1976-45-01, REI/1/5202-01-01, REI/1/5234-01-01, REI/1/4940-01-01, RGC/3/4816-01-01, and REI/1/0018-01-01. Xiaonan He was supported by the foundation of the National Natural Science Foundation of China (No. 62272327). \\

\noindent
\textbf{Competing Interests: }The authors have declared no competing interests.\\

\noindent
\textbf{Author Contribution Statements: }J.Z. and X.G. conceived of the presented idea. J.Z. designed the computational framework and analysed the data. J.Z, L.S., J.X., X.C., Y.C., L.Z., X.L., B.Z. and X.H. conducted the clinical evaluation. X.G. supervised the findings of this work. J.Z., X.H., L.S., J.X. and X.G. took the lead in writing the manuscript and supplementary information. All authors discussed the results and contributed to the final manuscript.\\

\noindent
\textbf{Data availability: }The data that support the findings of this study are divided into two groups: shared data and restricted data. Shared data include the SKINCON dataset and the Dermnet dataset. The SKINCON dataset can be accessed at \url{https://skincon-dataset.github.io/}. The Dermnet dataset can be accessed at \url{https://www.kaggle.com/datasets/shubhamgoel27/dermnet}. The restricted in-house skin disease images used in this study are not publicly available due to restrictions in the data-sharing agreement.\\

\noindent
\textbf{Code availability: }To promote academic exchanges, under the framework of data and privacy security, the code proposed by SkinGPT-4 is publicly available at \url{https://github.com/JoshuaChou2018/SkinGPT-4}. In the case of non-commercial use, researchers can sign the license provided in the above link and contact J.Z. or X.G. to access the latest non-commercial trained model weights.

{
\bibliographystyle{IEEEtran}
\bibliography{reg}
}

\end{document}